\documentclass{article}

\usepackage{cite}
\usepackage[utf8]{inputenc}

\title{A Dynamic Tree Architecture for Hierarchical On-Chain Asset Management}

\author{
    Mojtaba Eshghie\textsuperscript{1}, Gustav Andersson Kasche\textsuperscript{2} \\
    \texttt{\{eshghie, gustavak\}@kth.se} \\
    KTH Royal Institute of Technology, Stockholm, Sweden
}
\date{} 

\usepackage{graphicx}
\usepackage{todonotes}
\usepackage{algorithm} 
\usepackage{algpseudocode} 
\usepackage{amsthm}
\newtheorem{definition}{Definition}
\usepackage{float}

\usepackage{comment}

\begin{document}

\maketitle

\begin{abstract} In this paper, we introduce the \textsc{Sarv}, a non-monolithic blockchain-based data structure designed to represent hierarchical relationships between digitally representable components. \textsc{Sarv}\ serves as an underlying infrastructure for a wide range of applications requiring hierarchical data management, such as supply chain tracking, asset management, and circular economy implementations. Our approach leverages a tree-based data structure to accurately reflect products and their sub-components, enabling functionalities such as modification, disassembly, borrowing, and refurbishment, mirroring real-world operations. The hierarchy within \textsc{Sarv}\ is embedded in the on-chain data structure through a smart contract-based design, utilizing Algorand Standard Assets (ASAs). The uniqueness of \textsc{Sarv}\ lies in its role-based and compact  architecture, its mutability, and a two-layer action authorization scheme that enhances security and delegation of asset management. We argue that \textsc{Sarv}\ addresses real-world requirements by providing a scalable, mutable, and secure solution for managing hierarchical data on the blockchain. \end{abstract}

\section{Introduction}

The increasing emphasis on sustainability and resource efficiency has led to the rise of the circular economy paradigm, which promotes the reuse, refurbishment, and recycling of products and their components to minimize waste and resource consumption \cite{morselettoTargetsCircularEconomy2020}. Implementing a circular economy requires robust systems for tracking products and their constituent parts throughout multiple life cycles. Blockchain technology has been recognized as a promising solution for providing transparency, traceability, and security in supply chain management \cite{circlechain}.

However, existing blockchain-based solutions often lack the flexibility and scalability needed to accurately represent hierarchical relationships between products and their sub-components, especially in contexts where products are frequently modified, disassembled, or reassembled \cite{heinesTokenizationEverythingFramework2021}. Traditional monolithic blockchain architectures are not well-suited for modeling complex, mutable hierarchies due to their rigidity and inefficiency in handling dynamic structures \cite{capturedcr}.

To address these challenges, we propose \textsc{Sarv}, a non-monolithic, blockchain-based data structure designed to represent and manage hierarchical relationships between digitally representable components. \textsc{Sarv}\ leverages a tree-based architecture that mirrors the physical composition of products and their sub-products, enabling accurate reflection of real-world operations such as disassembly, modification, and reuse. The hierarchy within \textsc{Sarv}\ is maintained on-chain through a smart contract-based design utilizing Algorand Standard Assets (ASAs) \cite{AlgorandStandardAssets}.

Our approach draws inspiration from the scalability and hierarchical nature of the Domain Name System (DNS) \cite{wesselsHowChromiumReduced2021}, aiming to provide a scalable hierarchical distributed system capable of handling complex structures efficiently. By adopting a similar hierarchical model, \textsc{Sarv}\ can manage large-scale hierarchies with improved scalability and performance.

Furthermore, we recognize that any blockchain-based solution intended for managing products and their components through multiple life cycles must meet several critical requirements. Specifically, it must support a \textit{mutable structure} to accommodate constant change, remain \textit{non-monolithic} to prevent systemic risk and enable flexible scaling, provide a \textit{managing authority} system that allows for decentralized yet controllable modifications, incorporate \textit{delegates} who can act on behalf of owners, maintain a \textit{product passport} ensuring detailed information retention, and achieve \textit{scalability} to handle extensive, evolving product hierarchies. Together, these requirements form the foundation of our design, guiding the development of \textsc{Sarv}’s architecture and its corresponding security and control mechanisms.

Our approach introduces a two-layer action authorization scheme that enhances security and control over the token hierarchy. This scheme allows for granular management of product tokens and their sub-tokens, supporting operations such as delegation and reassembly while ensuring proper authorization at each level of the hierarchy. By meeting these outlined requirements, \textsc{Sarv}\ moves beyond the limitations of existing blockchain structures, offering a practical and robust way to represent, manage, and evolve complex on-chain hierarchies.

In this paper, we detail the design and implementation of \textsc{Sarv}, demonstrate how it addresses the limitations of existing solutions, and discuss its applicability to real-world scenarios, particularly in the context of the circular economy. We also compare \textsc{Sarv}\ with related work in blockchain-based hierarchical data structures and supply chain management systems.

\section{Related Work}
Blockchain technology has found applications in various areas like supply chain management, decentralized finance (DeFi)~\cite{capturedcr,zhouSoKDecentralizedFinance2023c,eshghieCreationExploitationOracle2024b}, and identity systems. However, its use for managing complex, hierarchical data structures, especially in dynamic systems, is still underdeveloped. Current solutions face challenges in scalability, flexibility, and maintaining efficient structures for real-world applications.

Wei et al. surveyed blockchain data management systems, identifying issues with traditional architectures like standard and hybrid blockchains, which struggle with scalability and adaptability for hierarchical data. Although Directed Acyclic Graph (DAG)-based blockchains offer better throughput, they do not fully address the challenges of organizing and managing hierarchical relationships in dynamic systems \cite{wei2022survey}.

Tokenization has been proposed as a method to bridge blockchain with real-world applications. CircleChain, introduced by Eshghie et al., focuses on tokenizing second-life components in circular economies, using a role-based approach to enhance lifecycle traceability. However, its design is limited to linear token relationships, making it unsuitable for modeling complex, mutable hierarchies \cite{circlechain}. Schubert et al. developed a token design framework to standardize token systems and simplify their development. While helpful for standalone token use cases, this framework does not address the need for managing interdependent hierarchical structures \cite{schubertDevelopmentTokenDesign2021}.

Smart contracts have played a critical role in enabling decentralized control and automation of blockchain processes. Ahmed et al. explored blockchain-based identity systems, emphasizing the need for privacy, trust, and decentralized control. These findings highlight the potential of smart contracts to manage complex systems with fine-grained authorization, but their focus was on identity management, not hierarchical data \cite{ahmedBlockchainBasedIdentityManagement2022}. Tavakoli et al. combined blockchain with digital twins for predictive asset management, ensuring data reliability and provenance. However, their system was designed for relatively static setups and did not address dynamic modifications common in real-world hierarchical systems \cite{tavakoliBlockchainbasedDigitalTwin2023}.

The mentioned works fall short of providing solutions for dynamically evolving hierarchical systems. Most approaches are either too rigid or lack scalability for large-scale, mutable structures. Addressing these gaps requires a blockchain framework designed explicitly for managing dynamic hierarchies.
Our work builds on these efforts by introducing \textsc{Sarv}, a tree-based, non-monolithic architecture tailored for blockchain-based hierarchical data management. It combines the immutability and transparency of blockchain with the adaptability needed for real-world operations like disassembly, modification, and reassembly of components. \textsc{Sarv} addresses the limitations of existing approaches and provides a scalable, flexible solution for hierarchical systems.

\section{Solution}

\subsection{System Design}\label{section:ds}

\begin{definition}
  \textbf{Tree}: A tree is defined as a data structure consisting of one or multiple nodes connected without a cycle. The nodes in a tree are either tree root, internal nodes, or leaves. An internal node may have one or more child nodes and is called the parent of its children. Leaves have only a connection to their parent \cite{Tree}.
\end{definition}

\begin{definition}
  \textbf{Forest}: A forest is a collection of one or more trees. The trees in this collection need not be connected \cite{Tree}.
\end{definition}

\begin{figure}[H]
    \centering
    \includegraphics[width=1\textwidth]{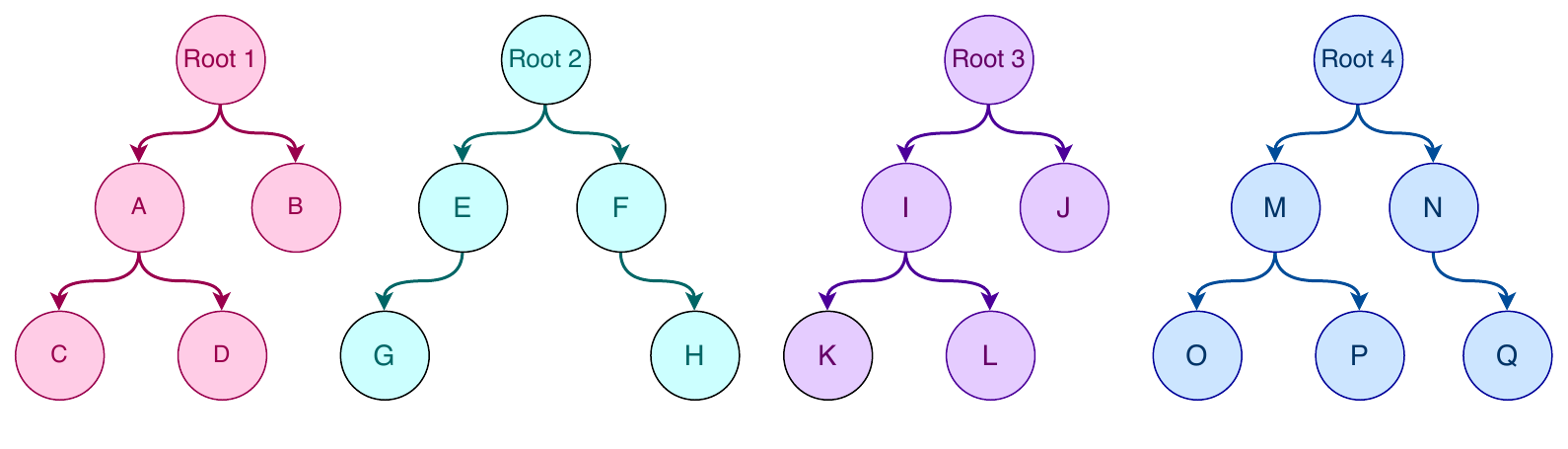}
    \caption{Simple forest of token trees denoting different products, their product parts, and the hierarchy of the physical product.}
    \label{fig:simple-token-tree}
\end{figure}

Based on the definition mentioned above, \textsc{Sarv}\ comprises trees (built by tokens, hence, token trees) that can change shape and reform the forest. A visualization of \textsc{Sarv}\ is depicted in Figure \ref{fig:simple-token-tree}. The bonds in this shape are shown using an arrow from parent to child. This bond corresponds to the relationship between product parts of the respective product.

\begin{figure}[H]
    \centering
    \includegraphics[width=1\textwidth]{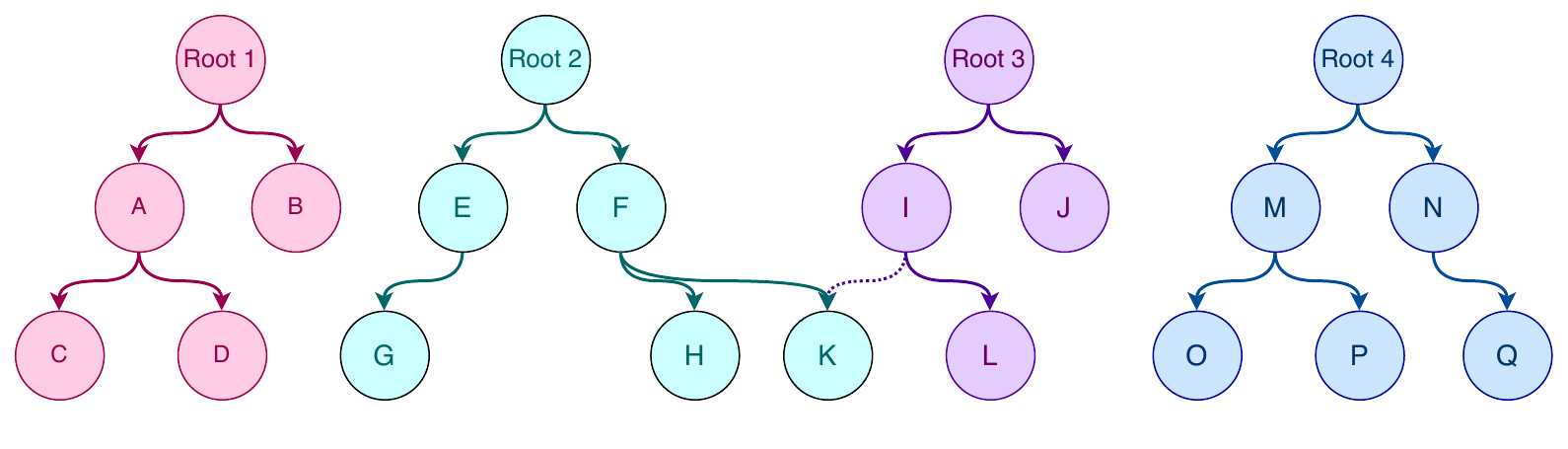}
    \caption{Breaking the bonds between one product part K and its original product token (with root 3) and build a new bond to another token tree (with root 2) under sub-product F.}
    \label{fig:breaking-joining-bonds}
\end{figure}

Operations on \textsc{Sarv}\ are defined to reflect real-world product management requirements in a circular supply chain. Therefore, reshaping and reforming the trees in the forest correspond to disassembling, assembling, refurbishing, or disposing of products in the real world.
Figure \ref{fig:breaking-joining-bonds} shows how a part (node K) from a product (root 3) can be separated and become part of another product (root 2).

\begin{figure}[H]
    \centering
    \includegraphics[width=0.7\textwidth]{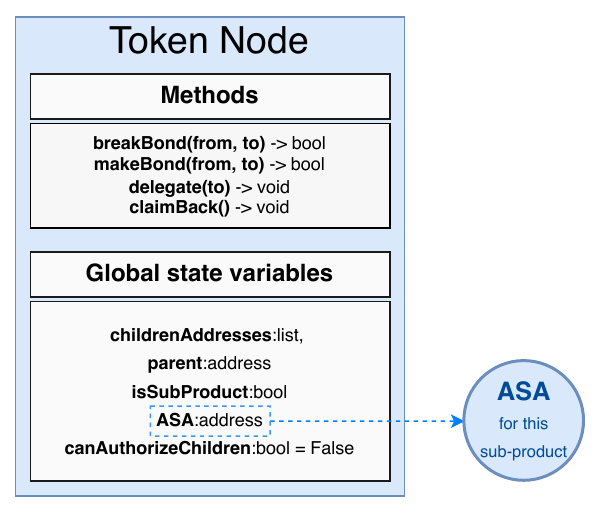}
    \caption{High-level view of each node of the tree in the system.}
    \label{fig:each-node}
    \centering
\end{figure}

Each node in Figure \ref{fig:breaking-joining-bonds} consists of an Algorand smart contract \cite{SmartContractDetails}, and can also be associated with a particular ASA. It is worth mentioning that the root of each tree can identify stand-alone products. In this order, if a user on Algorand blockchain holds the ASA related to a tree root, it means that the user is the owner of the associated product. Figure \ref{fig:each-node} shows the parts of a typical tree node. Since each node is a smart contract, it can have logic. This logic is used for reshaping the tree structure.

\subsubsection{The Two Layer Authorization Scheme}
\begin{figure}[H]
    \centering
    \includegraphics[width=1\textwidth]{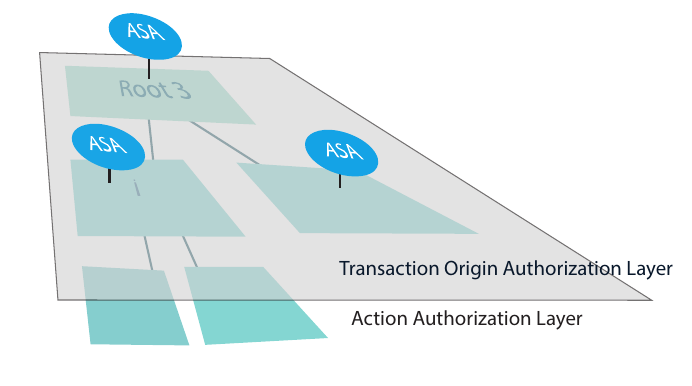}
    \caption{Layered authorization system}
    \label{fig:layered-arch}
\end{figure}

A combination of ASA and contract logic empowers creating of a two-layer authorization scheme for \textsc{Sarv}. As it is depicted in Figure \ref{fig:layered-arch}, one layer is the origin authorization which works by checking the owner of the ASA making the call. This layer checks if the transaction initiator is a blockchain user with the authority to perform the call. The second layer is related to the tree structure itself.
The general rule in the authorization of activities is that a parent \textit{may} be able to authorize the reformation of its children (and accordingly, the subtree that the children are a parent to).
Therefore, the tree root is the most authorized node meaning that it can change any \textit{bond} in the system. Besides the root of the token tree, any other internal (non-leaf) nodes can also manipulate their children given that any of their parents in the tree enables them to do so by setting the \textit{canAuthorizeChildren} (Figure \ref{fig:each-node}) to \textit{True}.

\subsubsection{Breaking a Bond}
\begin{figure}[H]
    \centering
    \includegraphics[width=1\textwidth]{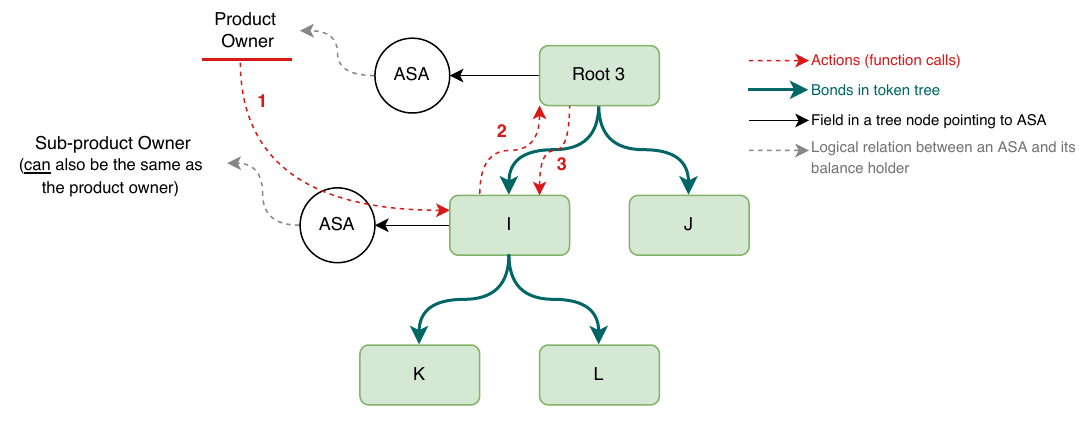}
    \caption{Break bond action (breaking mutual bond: parent to child and vice versa)}
    \label{fig:breakparentbond}
    \centering
\end{figure}

Breaking a bond should always be initiated from a parent node ASA owner to the child node. Figure \ref{fig:breakparentbond} demonstrates how bond-breaking works. Initially, the transaction is issued by the ASA holder of Root 3 (the parent). This call is shown by the dotted red line 1. Then, dotted red line 2 shows the next call in the chain of calls that asks the parent to remove its bond with the child. After the result of that call returns to the child I (red dotted line 3), I will also set its parent field to "NULL". If breaking of a bond is not followed by a make bond, the newly separated child's fields can be later changed to mark it as the root of a new tree. This is not enforced on-chain as \textsc{Sarv}\ focuses on offering a data structure and its primary tree management operations.

The pseudocode of the \textit{breakBond} function logic is demonstrated in Listing \ref{alg:breakbond}. The operation of the three arrows in Figure \ref{fig:breakparentbond} is performed in a distributed way on both the parent and child side. Lines $2$-$5$ of Listing \ref{alg:breakbond} match arrow 3 in Figure \ref{fig:breakparentbond}. Root 3 after breaking the bond will return \textit{True} to the caller (here, token node I) which enables it to break its bond with the parent. This is also demonstrated in lines $6$-$12$ in Listing \ref{alg:breakbond}. The \textit{call origin} (in line $2$) points to the immediate caller of the function.

\begin{algorithm}[H]
	\caption{Break a bond function}
	\begin{algorithmic}[1]
	
	\Function{breakBond}{from, to}
	\If{(call origin == any immediate child) AND (from == call origin AND to == this.address)}
    	\State this.childrenAddresses.remove(from)
    	\State \Return True
	\EndIf
	
	\If{(call origin == ASA owner of any parent with sub-tree authorization ability) OR (call origin == tree root)}
	\State result = call breakBond(this.address, parent) on this.parent
    	\If{result == True}
    	\State this.parent = NULL
    	\State \Return True
    	\EndIf
	
	\EndIf
	
	\State \Return False
	
	\EndFunction
	\end{algorithmic} 
	\label{alg:breakbond}
\end{algorithm} 

\subsubsection{Making a Bond}\label{subsection:makebond}
Making a bond can be translated into two following operations order:
\begin{enumerate}
    \item Adding a child to the parent in the child.
    \item Changing parent from NULL to a certain address.
\end{enumerate}
Furthermore, if the new child is part of another token tree (the parent is not NULL yet), before making a bond, that already existing bond should be broken. This is ensured in lines $9$-$12$ in Listing \ref{alg:makebond}. The first operation of the above list is performed in lines $3$-$8$ where apart from authorizing the call origin user, cycle creation is also avoided (\textit{if} condition in line $4$). The second operation is performed through lines $9$-$15$ in Listing \ref{alg:makebond}. This is done by first calling \textit{makeBond} function in the parent and seeing if the parent accepts this token node as a child. If it does (lines $11$-$13$), the parent field will also be changed in the child.

\begin{algorithm}[H]
	\caption{Make a bond function}
	\begin{algorithmic}[1]
    \Function{makeBond}{from, to}
    
    \State isFromAbove:bool = ((call origin == ASA owner of any parent) AND (parent.canAuthorizeChildren == True)) OR (call origin == tree root)
    
	\If{((call origin == ASA owner of the current node) AND (this.canAuthorizeChildren == True)) OR isFromAbove}
    	\If{(from == this.address) AND (to != this.parent) AND (to is not in this.childrenAddresses)}
    	    \State this.childrenAddresses.append(to)
            \State \Return True
        \EndIf
	\EndIf

    \If{(from == this.address) AND (this.parent == NULL) AND (ASA owner of the current node is the same as ASA owner of the `to`)}
        \State result = call makeBond(to, this.address) on `to` node 
        \If{result == True}
        \State this.parent = to
        \EndIf
        \State \Return True

	\EndIf

	\State \Return False

	\EndFunction
	\end{algorithmic} 
	\label{alg:makebond}
\end{algorithm} 

\subsubsection{Delegation}
We define delegation as an organization or user on blockchain being authorized to change the structure of the token tree (or part of it). This ability, however, should be temporary and the actual owner of the product or product part should be able to take control back whenever they wish. \\
Delegation in \textsc{Sarv}\ is done by transferring the ASA of that product or product part to another user (delegate address), and at the same time setting the \textit{isDelegated} field in the root (or sub-product) to True. The mentioned actions are done in lines $3$-$5$ in Listing \ref{alg:delegate}.
One important design choice is that you cannot explicitly delegate the root node. The reason is that taking back the control of a fully delegated tree would not be possible because of the \textit{claimBack} mechanism. The workaround is to have a child below the root and delegate that child instead of the root.

\begin{algorithm}[H]
	\caption{Delegation}
	\begin{algorithmic}[1]
	\Function{delegate}{to}
    	\State isFromAboveNotRoot:bool = (call origin == ASA owner of any parent) AND (node pertaining to the owner ASA.canAuthorizeChildren == True) AND (call origin != tree root)
        \If{isFromAboveNotRoot}
            \State this.isDelegated = True
            \State transfer(this.ASA, to)
        \EndIf
	\EndFunction
	\end{algorithmic}
	\label{alg:delegate}
\end{algorithm} 

\subsubsection{Claiming a Delegated Token Back}\label{subsection:takeback}
Taking back the control of the delegated sub-tree is easy since the control of any sub-tree is through its parent(s), therefore, the ASA holder of parent(s) can request to take the delegated child node back at any time they want. This is demonstrated in Listing \ref{alg:claimBack}.

\begin{algorithm}[H]
	\caption{Taking control back from the delegate}
	\begin{algorithmic}[1]
	\Function{claimBack}{}
	\State isFromAbove:bool = ((call origin == ASA owner of any parent) AND (parent.canAuthorizeChildren == True)) OR (call origin == tree root)
    \If{isFromAbove}
        \State this.isDelegated = False
        \State transfer(this.ASA, this.owner)
    \EndIf
	\EndFunction
	\end{algorithmic}
	\label{alg:claimBack}
\end{algorithm}

\subsection{Organizational Integration}

This section will motivate and describe the integration of the proposed solution with external organizations. It will also briefly describe how the system could be monetized.

\subsubsection{API}\label{subsection:api}
\begin{figure}[H]
    \centering
    \includegraphics[width=1\textwidth]{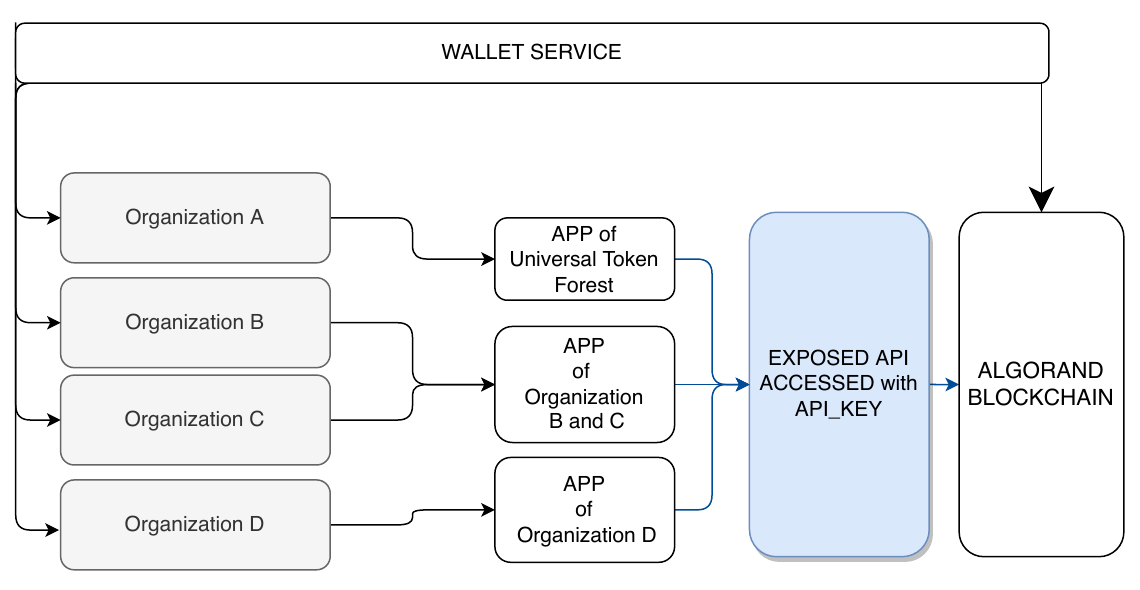}
    \caption{Example System architecture}
    \label{fig:architecture}
\end{figure}

The system architecture is defined to facilitate universal and simple use by organizations outside of the blockchain ecosystem without any engagement with blockchain-specific technology.
Therefore, the system core is an API where all the functions needed to build, mutate, and manage the data structure of \textsc{Sarv}\ on-chain are exposed. How the user interacts with our API is completely up to them as seen in Figure \ref{fig:architecture}. This gives the user the freedom to govern their use of \textsc{Sarv}\ as their industry sector demands. Various organizations may have further requirements that they can define on or off-chain. Our architecture gives the organization complete freedom over this.
When the organization invokes a function that requires on-chain authentication, it will use a wallet manager.
This greatly simplifies the UX of the system since no knowledge of Algorand or any blockchain is necessary to use \textsc{Sarv}. All the user needs to do is request the API for the \textsc{Sarv}\ actions they wish to perform, and the API will set up every transaction needed for \textsc{Sarv}\ on-chain and simply request the user for the sign using their private key in their wallet.

\subsubsection{API Monetizing}
As will be articulated in section \ref{section:discussion}, only basic tree-management operations are on-chain by the logic of the smart contracts. The more complex operations including keeping track of tokens can be done off-chain. Apart from a design choice to build a less complex system, this can also lead to a business model.  \\
The API is also used to monetize the system. API will be accessed using API keys associated with each organization. This gives us complete freedom in how we choose to monetize the off-chain services built on top of \textsc{Sarv}. We can bill the users for each API invocation or periodically for API access.

\section{Discussion}\label{section:discussion}
There might be specific technical challenges in the way of implementing this. An example is being forced to perform consecutive calls to functions in which case we might face limitations regarding maximum call depth.\\
There might be cases where we only need to read the data from the blockchain. In this case, it is possible to eliminate the problem by merely consecutively reading transactions and blocks to get the tree structure we want. Using this method instead of direct on-chain recursive calls eliminates the limitations related to the maximum depth of calls.

\subsection{Off-chain Operations}
With the minimalistic design of \textsc{Sarv}, it is possible to let more specific requirements of the system be defined and implemented by potential clients, organizations, or even as part of the API described in section \ref{subsection:api}. In the next subsections, we will go through some of these domain-specific requirements.
\subsubsection{Structural Violations Monitoring}
Operations such as ensuring a particular operation do not violate the token tree structure do not need to be done on-chain. Such operations can be done as part of the API (on-demand), periodically, or before issuing on-chain certain transactions such as makeBond which might add cycles to the tree.
\subsubsection{Product Passport}
Without having a proper way to retrieve product information and history, \textsc{Sarv}\ would not be very beneficial. There are two ways to integrate product (token) information and history into \textsc{Sarv}:
\begin{enumerate}
    \item Track smart ASAs of organizations, contract accounts, and transactions using an off-chain application. The whole life-long information of the token, hence the product will be available. Information includes origin, current situation, buyers, sellers, recyclers, and refurbishers. These can be added as a field to the smart contract of the root node of the product. Information regarding the product parts can also be added to the tree structure similarly based on the description of tree operations in section \ref{section:ds}.
    \item Another interesting feature of a blockchain-based system is that merely tracking the transactions that happen on the tokens, or the token trees reveal their buy, sell, and delegation history. This tracking is similar to what blockchain indexers do with more specialized functionality.
\end{enumerate}

\section{Conclusion}

We introduced \textsc{Sarv}, a novel non-monolithic blockchain-based data structure designed to represent and manage hierarchical relationships between digitally representable components. By leveraging a tree-based architecture and a two-layer action authorization scheme utilizing Algorand Standard Assets (ASAs) and smart contracts, \textsc{Sarv}\ provides a scalable, mutable, and secure solution for on-chain hierarchical data management.

Our approach addresses the limitations of existing blockchain solutions by enabling accurate reflection of real-world operations such as disassembly, modification, and reuse of products and their sub-components. \textsc{Sarv}'s design supports granular control over token hierarchies, facilitating operations like delegation and reassembly while ensuring proper authorization at each level.

We have demonstrated how \textsc{Sarv}\ fulfills the real-world requirements of applications within the circular economy and beyond, providing a practical implementation for tracking and managing products throughout multiple life cycles. Future work may explore the integration of \textsc{Sarv}\ with other blockchain platforms, performance optimization, and the development of standardized protocols for broader industry adoption.

\bibliographystyle{IEEEtran}
\bibliography{refs}

\end{document}